\documentclass[prc,amsmath,amssymb,twocolumn,preprintnumbers,longbibliography]{revtex4-2}
\usepackage{amsmath, amsthm, amssymb, braket}

\usepackage{amsfonts}

\usepackage{hyperref}
\usepackage[dvipdfmx]{graphicx}
\usepackage{color}
\usepackage{dcolumn}
\usepackage{bm}

\newcommand{\be}{\begin{equation}}
\newcommand{\ee}{\end{equation}}
\newcommand{\dd}{\mathrm{d}}
\newcommand{\br}{\boldsymbol{r}}
\newcommand{\bp}{\boldsymbol{p}}

\begin{document}
\allowdisplaybreaks[1]

\title{Spin-triplet proton--neutron pair in spin-dipole excitations
}

\author{Kenichi Yoshida}
\email[E-mail: ]{kyoshida@ruby.scphys.kyoto-u.ac.jp}
\affiliation{Department of Physics, Kyoto University, Kyoto, 606-8502, Japan}

\author{Yusuke Tanimura}
\affiliation{Department of Physics, Tohoku University, Sendai, 980-8578, Japan \\
Graduate Program on Physics for the Universe, Tohoku University, Sendai 980-8578, Japan}

\preprint{KUNS-2870}
\date{\today}

\begin{abstract}
\begin{description}
\item[Background] Spin-triplet ($S=1$) proton--neutron (pn) pairing in nuclei has been under debate. 
It is well known that 
the dynamical pairing affects the 
nuclear matrix element of the Gamow--Teller (GT) transition and the double beta decay. 
\item[Purpose] We investigate the effect of the pn-pair interaction in the $T=0, S=1$ channel 
on the low-lying spin-dipole (SD) transition. We then aim at 
clarifying the distinction of the role in between the SD and GT transitions.
\item[Method] We perform a three-body model calculation for the transition $^{80}\mathrm{Ni}\to{}^{80}\mathrm{Cu}$, 
where $^{78}$Ni is taken as a core. 
The strength of the pair interaction is varied to see the effect on the SD transition-strength distribution. 
To fortify the finding obtained by the three-body model, we employ the nuclear energy-density functional method for 
the SD transitions in several nuclei, where one can expect a strong effect. 
\item[Results] The effect of the $S=1$ pn-pair interaction depends on the spatial overlap of the pn pair 
and the angular momentum of the valence nucleons;
the higher the angular momentum of the orbitals, the more significant the effect. 
\item[Conclusions]
The dynamical $S=1$ pairing is effective even for SD states 
although the spatial overlap of the pn pair can be smaller than GT states. 
The SD transition involving high-$\ell$ orbitals with the same principal quantum number is 
strongly affected by the dynamical $S=1$ pairing. 
\end{description}
\end{abstract}

\maketitle

\section{Introduction}\label{intro}
Pairing is a ubiquitous many-body correlation emerging in various systems, including a nuclear system~\cite{bri05}.
The superfluidity or superconductivity 
is mostly understood by the spin-singlet pairing~\cite{sch64}. 
New facets of the nuclear pairing
 show up as a spatially-correlated two-neutron 
in dilute and weak-binding systems~\cite{mig72,ber91,zhu93,deb97,mat04,mat06,hag05,pil07}. 
The spatial localization of two neutrons has been confirmed experimentally~\cite{nak06,kub20}.
Meanwhile, there has been an enduring discussion on another type of exotic unconventional pairing in nuclei: 
the $^3S_1$ correlation of isoscalar (IS) proton--neutron (pn) pairs in $N \sim Z$ nuclei~\cite{fra14}.
The emergence of the $S=1$ pn-pair condensation is still controversial.

The fluctuation of the $S=1$ pair field and its effect on the observables 
has been discussed recently by investigating, {\it e.g.}, 
the pn-pair transfer-type modes of excitation~\cite{yos14,lit18,cha21} 
and the spin susceptibility~\cite{yos21}. 
Furthermore, it has been known that the dynamical $S=1$ pairing 
lowers the Gamow--Teller (GT) states in energy and thus shortens the 
$\beta$-decay half-lives of neutron-rich nuclei~\cite{eng99} including deformed nuclei~\cite{yos13}. 
The nuclear matrix elements of the double beta decay are also affected 
by the $S=1$ pn-pair interaction~\cite{eng88}. 

The effects of the $S=1$ pair interaction on the GT transition strengths have been thoroughly studied in $N = Z$ 
odd-odd nuclei with a three-body model of two nucleons around a spherical core~\cite{tan14,tan16,min18}. 
A remarkable feature found in $N=Z$ odd-odd nuclei with an $LS$-closed core ($^4$He, $^{16}$O, $^{40}$Ca) is 
the appearance of the low-energy state with a strong GT strength. 
A similar finding is obtained by employing a nuclear energy-density functional (EDF) method~\cite{bai14}.
The low-energy GT states have been indeed identified experimentally 
in the transitions of $^{18}\text{O}\to{}^{18}\text{F}$~\cite{fuj19} 
and $^{42}\text{Ca}\to{}^{42}\text{Sc}$~\cite{fuj14,fuj15}.

The spin-dipole (SD) excitation is induced by the spin operator, similarly to the GT excitation.
The principal quantum numbers of the single-particle orbital 
of a proton and a neutron differ by one unit contrary to $\Delta N=0$ in the GT transition. 
The spatial overlap is thus weaker, and an effect of the pn-pair interaction has been overlooked. 
The high-energy first-forbidden $\beta$ decay occurs in neutron-rich nuclei 
due to the imbalanced Fermi levels of neutrons and protons~\cite{yos17}. 
To describe well the $\beta$-decay properties, needed is examining 
a role of the dynamical $S=1$ pairing in the forbidden transitions as well. 

Therefore, in this article, 
we are going to investigate 
the role of the $S=1$ pn-pair interaction in the SD transitions. 
To this end, 
we employ a three-body model to obtain an essential feature of the $S=1$ pairing 
in the SD excitations. 
To make what we find in the three-body model analysis solid and secure, 
we further perform the nuclear density-functional theory (DFT) calculation; 
a nuclear EDF method is utilized, which is a theoretical model being capable of 
handling nuclides with arbitrary mass numbers in a single framework~\cite{ben03,nak16}. 

This paper is organized in the following way: 
three-body model analysis is performed in Sec.~\ref{model}, 
where the low-lying SD $1^-$ state in $^{80}$Cu is studied;  
Sec.~\ref{Skyrme_SD} is devoted to the discussion 
by employing the nuclear EDF method 
to fortify the findings obtained by the three-body model calculation; 
then, a summary is given in Sec.~\ref{summary}.

\section{Three-body model analysis}\label{model}

In this section, we employ the three-body model to make a qualitative analysis on 
the effect of the $S=1$ pn-pair interaction on the low-lying SD $1^-$ states. 
Based on the analytic form of the matrix element of the $S=1$ interaction, 
we discuss the relation between the quantum numbers of single-particle states 
involved in the transition and the gain of energy due to the $S=1$ interaction. 

\subsection{Three-body model}
The details of the model are described in Refs.~\cite{ber91,tan14,hag05}. 
We thus briefly recapitulate the basic equations relevant to the present study. 
The Hamiltonian of the present model is given as
\be
H=\sum_{i=1,2}
\left[ \dfrac{\bp^2_i}{2m}+V_{Nc}(r_i)\right]+\frac{1}{2}\sum_{i \ne j}V_{NN}(\br_i,\br_j),
\label{Hami}
\ee
where $m$ is the mass of a nucleon, $r=|\br|$, and the recoil motion of the core is neglected. 
The potential between the nucleon ($N$) and the core nucleus ($c$), $V_{Nc}$, is given by 
\be
V_{Nc}(r)=\left[
V_0 + V_{ls}r_0^2(\boldsymbol{\ell} \cdot \boldsymbol{s})\dfrac{1}{r}\dfrac{\dd}{\dd r}\right]f(r)+V_C(r)
\ee
with $f(r)=[1+\exp(r-R)/a]^{-1}$ and the Coulomb potential $V_C(r)$ for a proton. 
The interaction between nucleons, $V_{NN}$, is given by 
\begin{align}
V_{nn}(\br,\br^\prime) =& v_s P_s[1-x_sf(r)]\delta(\br - \br^\prime), \\
V_{pn}(\br,\br^\prime) =& v_s P_s[1-x_sf(r)]\delta(\br - \br^\prime) \notag \\
&+ v_t P_t[1-x_tf(r)]\delta(\br - \br^\prime) \label{eq:spin_trip_int}
\end{align}
for neutron--neutron and proton--neutron, respectively, 
where $P_s$ and $P_t$ is a projector onto the $S=0$ and $S=1$ two-nucleon state, respectively.  

\subsection{Numerical procedures}
The parameters of the Woods--Saxon potential are the standard one given in p.239 of Ref.~\cite{BM1}. 
The Coulomb potential for a proton is obtained for the uniform charge distribution with the radius $R$. 
The single-particle states in the Woods--Saxon potential 
are obtained in a spherical box of 30 fm. 
The continuum states are then discretized. 
When diagonalizing the Hamiltonian (\ref{Hami}), 
the single-particle states are truncated with a cut-off energy at 10 MeV, which is enough to cover 
a few major shells above the Fermi level that are relevant to low-lying states. 
The strength of the $S=0$ pair interaction $v_s$ is set as $-500$ MeV fm$^3$ 
for the qualitative discussion. 
We note that, with the present set up, the ground-state binding energy 
of the three-body system $^{78}{\rm Ni} + 2n$ is 4.0 MeV, which is comparable to the estimated value of 4.5 MeV~\cite{Huang_2021,Wang_2021}. 
To see the effect of the $S=1$ pair interaction, $v_t$ is varied 
by multiplying a factor $f_{NN}$ as $v_t=f_{NN}\times v_s$.
The parameters $x_s$ and $x_t$ are set as 0.5, corresponding to the so-called mixed-type pair interaction.

\begin{figure}[t]
\includegraphics[width=8.5cm]{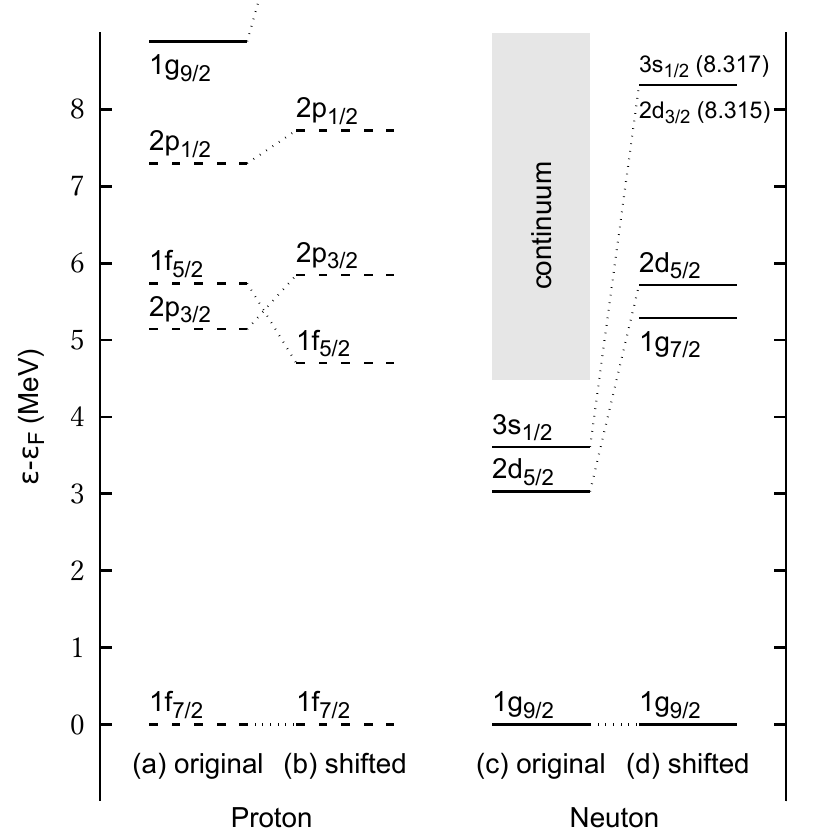}
\caption{\label{fig:sp_energy} 
Single-particle energies of the valence nucleons relative to the Fermi levels of the $^{78}$Ni core.
Shown are the proton single-particle
states with the (a) original and (b) shifted ($-40$ MeV)  Woods-Saxon potentials, respectively, 
while the neutron single-particle states are depicted in (c) and (d). 
Positive- and negative-parity states are shown by solid and dashed lines, respectively. 
The shaded area represents the continuum region of neutron states. 
Notice that the $2d_{3/2}$ and $3s_{1/2}$ states in case (d) are almost degenerate, 
which are located at 8.315 and 8.317 MeV, respectively. }
\end{figure}

Shown in Fig.~\ref{fig:sp_energy} are 
the single-particle states near the Fermi levels relevant to the discussion below. 

\subsection{Low-lying SD states in $^{80}$Cu}\label{Ni_SD}

We consider the response to 
the SD operator in the $(p,n)$ channel defined by
\begin{align}
	F^J_K
	=&\dfrac{1}{\sqrt{2}} \sum_{ss^\prime tt^\prime}\int \dd \br r
	\psi^\dagger(\br s^\prime t^\prime)\psi(\br s t) \times \notag \\
	&\hspace{1.5cm} \langle s^\prime|[Y_1 \otimes \boldsymbol{\sigma}]^J_K |s\rangle\langle t^\prime|\tau_{-1}|t \rangle, \label{SD_op}
\end{align}
where $\boldsymbol{\sigma}$ and $\tau=(\tau_{+1},\tau_0,\tau_{-1})$ denote the spherical components of the Pauli matrix of spin and isospin, and $\psi^\dagger(\br s t), \psi(\br s t)$ represent the nucleon field operators. 
The reduced transition probability
is given as 
\be
	B(SD,J; J_i \to J_f)=\frac{1}{2J_i +1} | \langle f ||F^J || i \rangle |^2.
\ee

\begin{figure}[t]
\begin{center}
\includegraphics[scale=0.39]{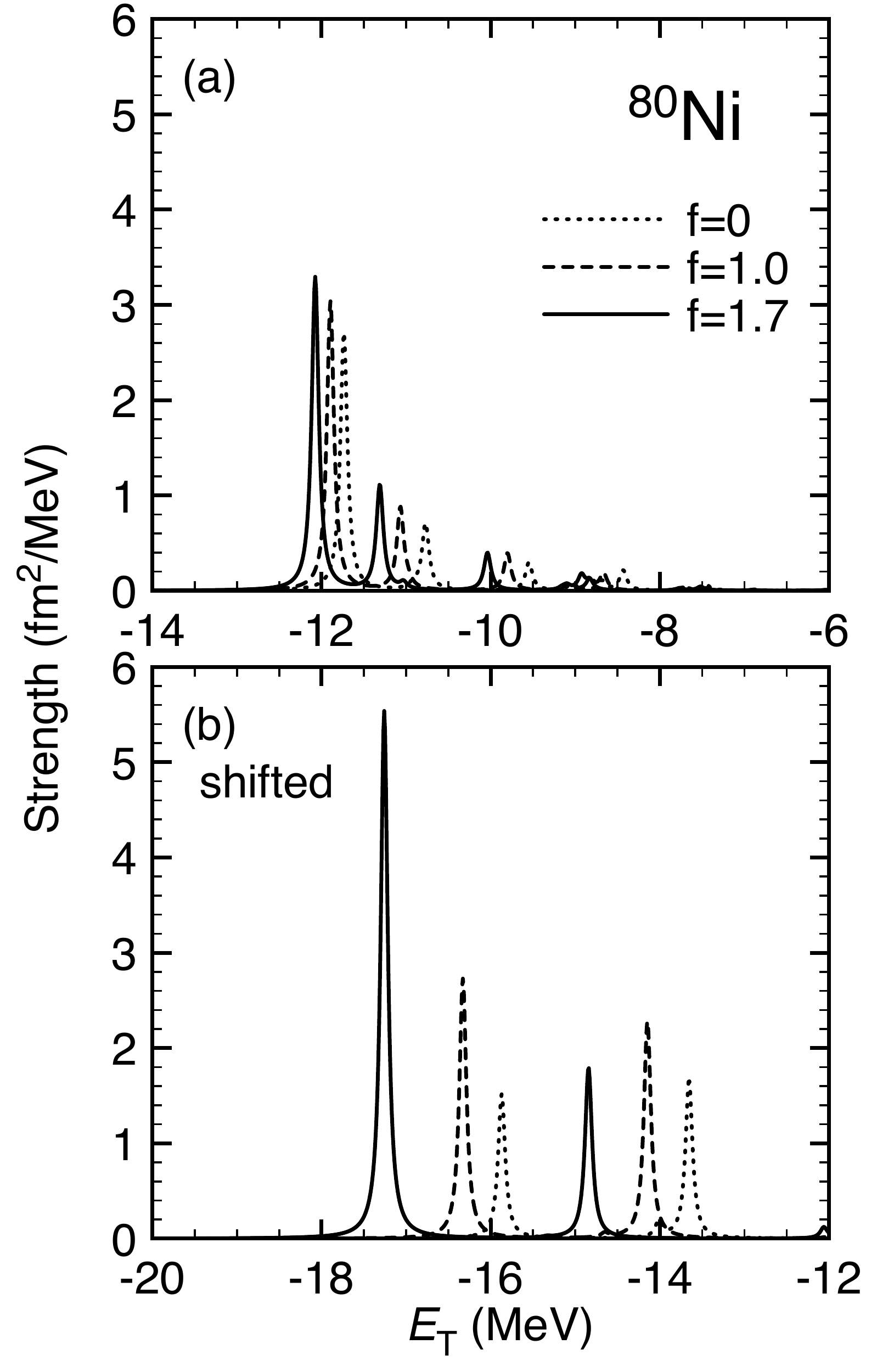}
\caption{\label{fig:SD1_strength} 
(a) Calculated distributions for the SD $1^-$
transition as functions of the excitation energy with respect to the target nucleus. 
The smearing parameter $\gamma=0.1$ MeV is used.  
The result obtained with $f_{NN}=0,\ 1.0$, and $1.7$ is depicted by the dotted, dashed, and solid lines, respectively. 
(b) Results obtained by changing the depth of the WS potential by $-40$ MeV. 
}
\end{center}
\end{figure}

We show in Fig.~\ref{fig:SD1_strength}(a) the calculated SD $J=1$ transition strengths 
in ${}^{80}\mathrm{Ni} \to {}^{80}\mathrm{Cu}$. 
One can see a prominent peak around $E_T=-12$ MeV and several states with a small transition strength. 
The prominent state corresponds to 
the low-lying $1^-$ state appearing at $-11.5$ MeV
in the self-consistent calculation based on the nuclear EDF method~\cite{yos19}; see Fig.~5(b) there. 
This correspondence shows that the present three-body model describes well 
the low-lying states in a nucleus with a spherical core plus two valence nucleons. 
It is noted that the $1^-$ state around $-12$ MeV plays a crucial role in the isotopic dependence of the $\beta$-decay half-lives in the Ni isotopes. 
We are thus going to look into the microscopic structure of this $1^-$ state. 

The effect of the $S=1$ pair interaction is investigated. 
To this end, we vary the strength of the interaction. 
The result obtained with $f_{NN}=0, 1.0,$ and $1.7$ is depicted by 
the dotted, dashed, and solid line, respectively, in the figure. 
The role of the $S=1$ pair interaction depends on the state; 
the low-lying state is weakly affected by the interaction compared with the 
second peak. 
Looking into the details of these states, 
we try to understand the role of the $S=1$ pair interaction 
in the SD excitations. 

The lowest and second peaks are constructed mainly by the 
$\nu 2d_{5/2} \otimes \pi 2p_{3/2}$ and $\nu 2d_{5/2} \otimes \pi 1f_{5/2}$ configurations, respectively. 
The shift in the energy and the enhancement in the transition strength
are governed by the two-body matrix element: 
\begin{align}
	&\langle a^\prime b^\prime,J|V_{S=1}|ab,J\rangle =R_{a^\prime b^\prime ab} \times A_{a^\prime b^\prime ab}^J, \\
	&R=v_t\int \dd r \dfrac{1}{r^2}[1-x_t f(r)]u_{a^\prime}^*(r )u_{b^\prime}^*(r) u_a(r) u_b(r),\\
	&A^J=\dfrac{1}{8\pi} \hat{j}_{a^\prime} \hat{j}_{b^\prime} \hat{j}_a \hat{j}_b
	\left[  
	\begin{pmatrix}
	j_{a^\prime} & j_{b^\prime} & J \\
	\frac{1}{2} & \frac{1}{2} & -1
	\end{pmatrix}
	\begin{pmatrix}
	j_a & j_b & J \\
	\frac{1}{2} & \frac{1}{2} & -1
	\end{pmatrix}
	\right. \notag \\
	&+\delta_{\ell_{a^\prime}+\ell_{b^\prime}+J,\text{odd}}\delta_{\ell_a+\ell_b+J,\text{odd}}
	(-1)^{\ell_{a^\prime}+\ell_a+j_{b^\prime}+j_b}\times \notag \\
	&
	\left. \hspace{3cm}
	\begin{pmatrix}
	j_{a^\prime} & j_{b^\prime} & J \\
	\frac{1}{2} & -\frac{1}{2} & 0
	\end{pmatrix}
	\begin{pmatrix}
	j_a & j_b & J \\
	\frac{1}{2} & -\frac{1}{2} & 0
	\end{pmatrix}
	\right], 	\label{eq:angular}
\end{align}
where $\hat{j}=\sqrt{2j+1}$, and $u_a(r)$ is the radial wave function of the single-particle orbital.
The diagonal matrix element $(R \times A^J)$ of the $\nu 2d_{5/2} \otimes \pi 2p_{3/2}$ and $\nu 2d_{5/2} \otimes \pi 1f_{5/2}$ configurations 
is $-5.60 \times 0.05 =-0.27$ MeV and $-3.60 \times 0.12 = -0.44$ MeV, respectively in the case of $f_{NN}=1.7$. 
The difference in the diagonal matrix element accounts for 
a stronger effect of the $S=1$ pair interaction seen on the second peak.
The angular part $A^J$ of the diagonal matrix element for $J^\pi=1^-$ reads
\begin{subequations}
\begin{align}
A^{1^-}(\ell_{j_>},(\ell-1)_{j_>};\ell_{j_>},(\ell-1)_{j_>})&=\dfrac{1}{8\pi}\dfrac{\ell(\ell+1)}{2\ell+1}, \\
A^{1^-}(\ell_{j_<},(\ell-1)_{j_>};\ell_{j_<},(\ell-1)_{j_>})&=\dfrac{1}{2\pi}\dfrac{\ell^3}{(2\ell+1)(2\ell-1)}, \label{eq:SD1_max}\\
A^{1^-}(\ell_{j_<},(\ell-1)_{j_<};\ell_{j_<},(\ell-1)_{j_<})&=\dfrac{1}{8\pi}\dfrac{\ell(\ell-1)}{2\ell-1}, 
\end{align}
\label{eq:SD1}
\end{subequations}
where $\ell_{j_\gtrless}$ denotes the orbital angular momentum satisfying $j=\ell \pm \frac{1}{2}$. 
The proton--neutron configuration involving the $j_>$ and $j_<$ orbitals, Eq.~(\ref{eq:SD1_max}),  
have the largest matrix element as far as 
$\ell$ of the configurations is the same. 
Thus, we see a more substantial effect of the $S=1$ pair interaction for the $\nu 2d_{5/2} \otimes \pi 1f_{5/2}$ configuration 
though the spatial part $R$ is slightly smaller due to the difference in the number of nodes. 
Notice that the spatial part is larger in the GT excitation because the GT operator 
does not change the spatial structure, and the angular part is as large as the SD excitations as given in the Appendix, 
leading to a stronger effect of the $S=1$ pair interaction in the GT excitation.

Another feature seen in Eq.~(\ref{eq:SD1}) is that the configuration involving a high-$\ell$ orbital acquires 
a large matrix element. 
Thus, we can expect that 
the $S=1$ pair interaction significantly affects the configuration composed of the $\nu 1g_{7/2}$ orbital. 
We are going to investigate this in the following subsection. 

\subsection{Role of a high-$\ell$ orbital}\label{high-l}

With the present parameters, 
the $\nu 1g_{7/2}$ orbital is located far above the Fermi level and embedded in the continuum states. 
To see the role of this high-$\ell$ orbital in the SD excitations, 
we deepen the potential so that the $\nu 1g_{7/2}$ orbital appears near the Fermi level.
The single-particle levels with the shifted potential is also shown in Fig.~\ref{fig:sp_energy}. 

Figure~\ref{fig:SD1_strength}(b) shows the SD transition strengths. 
A prominent state in low energies is constructed mainly by the $\nu 1g_{7/2} \otimes \pi 1f_{5/2}$ configuration. 
As expected, we see a strong effect of the $S=1$ pair interaction. 
Since the $\nu 1g_{7/2}$ and $\pi 1f_{5/2}$ orbitals have the same number of nodes, 
the spatial part is also large. 
The second peak in Fig.~\ref{fig:SD1_strength}(b) is also strongly affected by the $S=1$ pair interaction. 
This state is generated predominantly by the $\nu 2d_{5/2}\otimes \pi 1f_{5/2}$ configuration, 
corresponding to the second peak in Fig.~\ref{fig:SD1_strength}(a). 

\section{Discussion using Nuclear DFT}\label{Skyrme_SD}
We have found that the collective shift in the low-lying SD states due to the 
residual interactions is mostly governed by the diagonal matrix element. 
The larger the spatial and angular parts of the matrix element, 
the more strongly the $S=1$ pair interaction affects the SD states. 
For a larger spatial overlap, 
the orbitals should have the same number of nodes. 
High-$\ell$ orbitals acquire a large matrix element in the angular part. 
Since the SD states are generated by not only the pair interactions 
but the spin--isospin interactions discarded in the model study above, 
we employ nuclear density-functional theory (DFT) as a realistic calculation 
and try to fortify the finding obtained by the three-body model. 
Here we extend our discussion to SD $0^-$ and $2^-$ states. 

\subsection{Nuclear EDF method}\label{Skyrme_model}
We perform a self-consistent Kohn--Sham--Bogoliubov (KSB) and the 
proton--neutron quasiparticle-random-phase approximation (pnQRPA) calculation. 
The details of the calculation scheme are found in Refs.~\cite{kas21} and \cite{yos13} 
for the KSB and pnQRPA, respectively. 
In brief, 
we solve the KSB equation in the coordinate space using cylindrical coordinates
$\boldsymbol{r}=(\rho,z,\phi)$ with a mesh size of
$\Delta\rho=\Delta z=0.6$ fm and a box
boundary condition at $(\rho_{\mathrm{max}},z_{\mathrm{max}})=(14.7, 14.4)$ fm. 
The quasiparticle (qp) states are truncated according to the qp 
energy cutoff at 60 MeV, and 
the qp states up to the magnetic quantum number $\Omega=23/2$
with positive and negative parities are included, with $\Omega$ being the $z$-component of the angular momentum. 
We introduce the truncation for the two-quasiparticle (2qp) configurations in the QRPA calculations,
in terms of the 2qp-energy as 60 MeV. 
For the normal (particle--hole) part of the EDF,
we employ the SGII functional~\cite{gia81}. 
For the pairing (particle--particle, p--p) energy, we adopt the one in Ref.~\cite{yam09}
that depends on both the IS and IV densities, 
in addition to the pair density, with the parameters given in
Table~III of Ref.~\cite{yam09}. 
The same pairing EDF is employed for the $S=0$ pn-pairing 
in the pnQRPA calculation, 
while the linear term in the IV density is dropped. 
The strength of the $S=1$ pn-pair interaction is varied by multiplying by a factor $f_{NN}$. 
The SD transition matrix elements are calculated as in Ref.~\cite{yos20}. 

\subsection{SD $0^-$ state in $^{132,134}$Nb}\label{Skyrme_cal0}
Among the SD excitations, the angular part of the diagonal matrix element 
is the largest in the $J=0$ transition; see Eq.~(\ref{eq:SD0}) in the Appendix. 
Therefore, we can expect a strong effect of the $S=1$ pair interaction to appear 
in the SD $0^-$ state involving such as $\nu 1g_{7/2} \otimes \pi 1f_{7/2}$ or 
$\nu 1h_{9/2} \otimes \pi 1g_{9/2}$ configuration. 
They correspond to the neutron-rich regions around Sc and Nb isotopes near the drip line. 
Since the neuron $1g_{7/2}$ orbital is located above the $2d_{5/2}$ orbital with the SGII functional,
$^{78}$Ca is a candidate for the study. However, we find it unbound. 
Therefore, we are going to investigate the neutron-rich Zr isotopes as an example.

\begin{figure}[t]
\begin{center}
\includegraphics[scale=0.27]{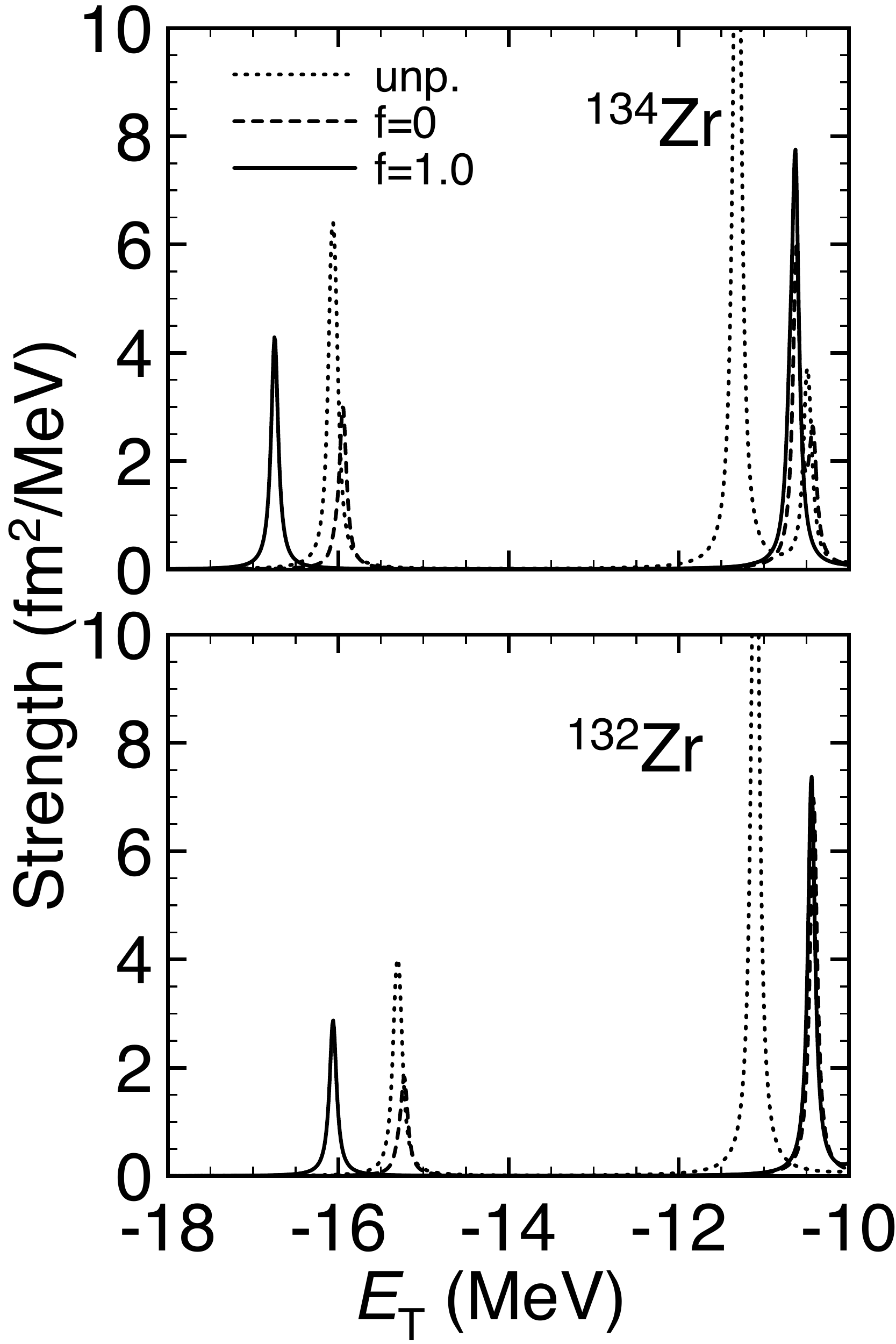}
\caption{\label{fig:SD0_strength} 
Similar to Fig.~\ref{fig:SD1_strength} but 
obtained by employing the Skyrme-KSB+pnQRPA with the SGII functional for the SD $0^-$ transition 
in $^{132,134}$Zr $\to ^{132,134}$Nb.  
The results obtained with $f_{NN}=0$ and $1.0$ are depicted by the dashed and solid lines, respectively. 
Shown is also the unperturbed strengths by the dotted line.
}
\end{center}
\end{figure}

The ground state of $^{132,134}$Zr 
is calculated to be spherical, though we have local minima at finite deformation. 
The neutron occupation probability of the $1h_{9/2}$ orbital is 0.04 and 0.06 in $^{132}$Zr and $^{134}$Zr, respectively.
Figure~\ref{fig:SD0_strength} shows the SD $0^-$ transition-strength distribution in the thus calculated $^{132,134}$Zr.
The low-energy peak is predominantly constructed by 
the $\nu 1h_{9/2} \otimes \pi 1g_{9/2}$ configuration.
Due to the repulsive character of the spin--isospin residual interactions, 
the 2qp excitation of $\nu 1h_{9/2} \otimes \pi 1g_{9/2}$ is shifted higher in energy when $f_{NN}=0$. 
However, 
a tiny difference between the QRPA energy and the unperturbed one 
indicates the weak collectivity of the low-lying state at $E_{\mathrm T}=-15.2 (-16.0)$ MeV in $^{132}$Zr ($^{134}$Zr). 
The transition strengths in a low-energy region are reduced 
due to the spin--isospin residual interactions, and they are brought into the giant resonance.
 
Since the pairing matrix elements entering into the QRPA equation 
are approximately proportional to $uuuu$ or $vvvv$ of the Bardeen--Cooper--Schrieffer amplitude~\cite{rin80}, 
the p--p type or hole--hole type excitation acquires a large matrix element.
Therefore, the low-lying SD $0^-$ state here is strongly affected by the dynamical pair interaction. 
Indeed, the $S=1$ pair interaction lowers the energy and enhances the strength of the SD $0^-$ state, 
as predicted based on the finding above. 
We can expect the $\beta$-decay rate in the Zr isotopes near the drip line is sensitively determined by 
the $S=1$ pair interaction through the first-forbidden transition 
because the positive-parity states, involving the $\pi 1h_{11/2}$ orbital, 
show up in relatively higher excitation energy. 
It should be noted that extracting the details of nuclear-structure information 
from the $\beta$-decay rate with such a high $Q_\beta$ value 
requires a careful treatment of the Coulomb potential~\cite{hor21}. 

\subsection{SD $2^-$ state in $^{120,122}$Sb and $^{68,70}$Cu}\label{Skrme_cal2}
The SD $2^-$ states appear in low energy widely in the nuclear chart~\cite{did94}. 
It has thus been investigated as a unique first-forbidden $\beta$ decay~\cite{hom96}, 
and is proposed as a probe of the physics beyond the Standard Model~\cite{gli17}. 
A neutron intruder orbital plays a decisive role in the occurrence of the low-lying $2^-$ state.
From Eq.~(\ref{eq:SD2}), we can expect a strong effect of 
the $S=1$ pair interaction for the configurations such as $\nu 1f_{7/2}\otimes \pi 1d_{3/2}$, 
$\nu 1g_{9/2}\otimes \pi 1f_{5/2}$, $\nu 1h_{11/2}\otimes \pi 1g_{7/2}$, and $\nu 1i_{13/2}\otimes \pi 1h_{9/2}$.

\begin{figure}[t]
\begin{center}
\includegraphics[scale=0.24]{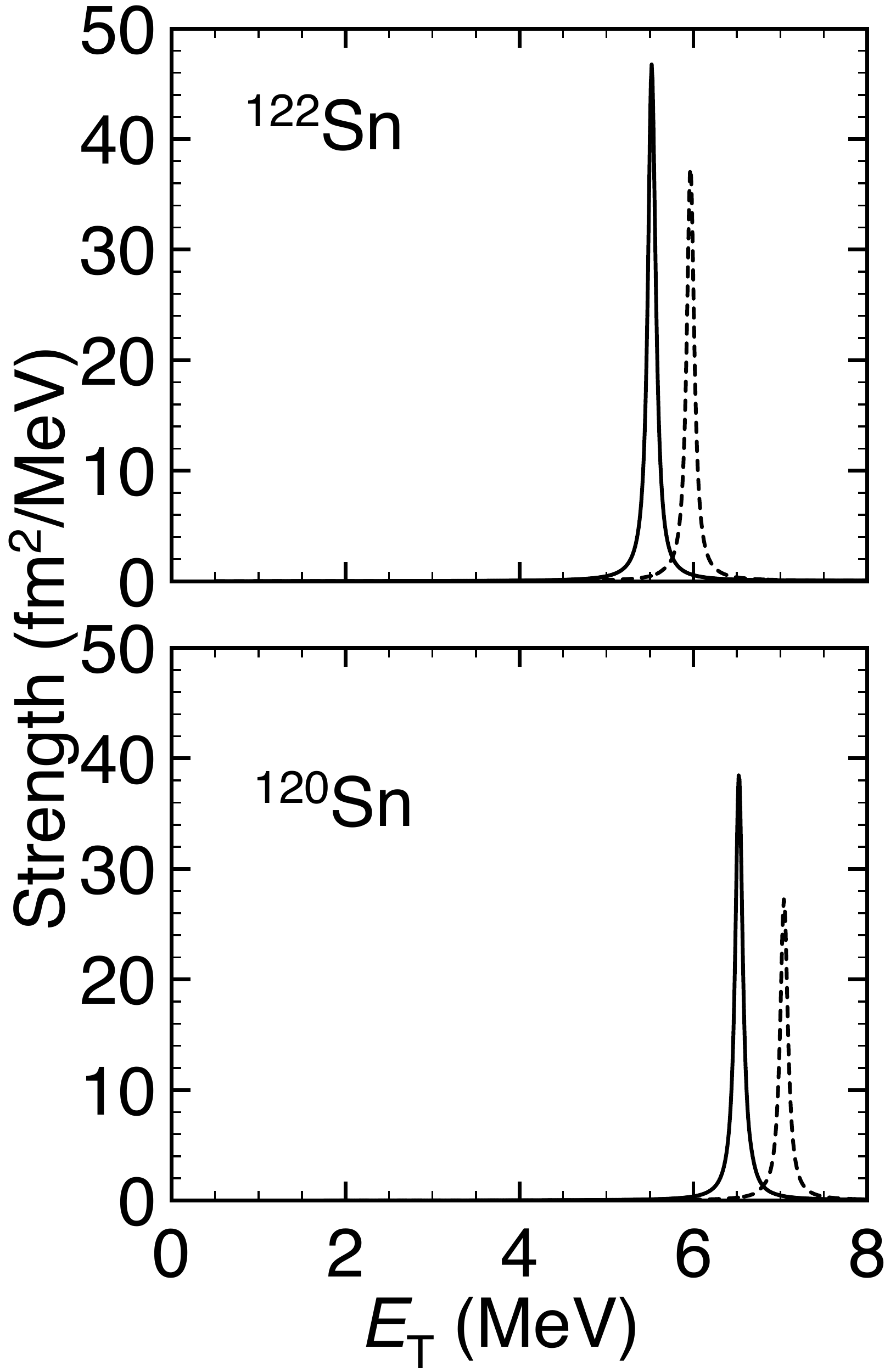}
\includegraphics[scale=0.24]{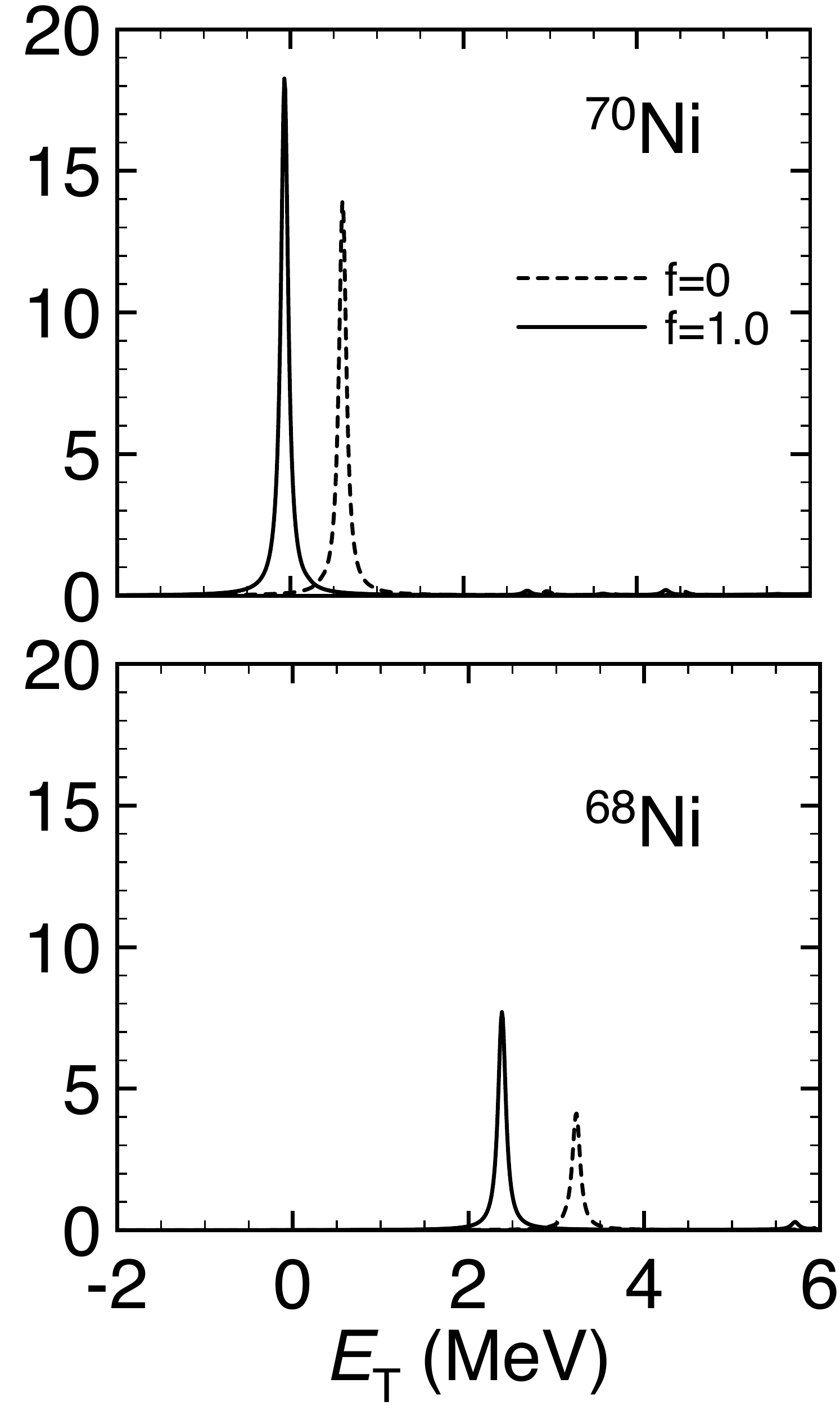}
\caption{\label{fig:SD2_strength} 
Same as Fig.~\ref{fig:SD0_strength} but 
for the SD $2^-$ transition 
in $^{120,122}$Sn $\to ^{120,122}$Sb and $^{68,70}$Ni $\to ^{68,70}$Cu.  
}
\end{center}
\end{figure}

Around $^{120}$Sn, the $\nu 1h_{11/2}$ and $\pi 1g_{7/2}$ orbitals are located 
near the Fermi level. 
The neutron occupation probability of the $1h_{11/2}$ orbital is calculated to be 
0.28 (0.37) in $^{120}$Sn ($^{122}$Sn).
Similarly, the $\nu 1g_{9/2}$ and $\pi 1f_{5/2}$ orbitals are placed near the Fermi level around $^{68}$Ni. 
The neutron occupation probability of the $1g_{9/2}$ orbital is 0.09 (0.23) in $^{68}$Ni ($^{70}$Ni) 
in the present calculation. 

Figure~\ref{fig:SD2_strength} shows the calculated transition-strength distribution 
in $^{120,122}$Sn and $^{68,70}$Ni. 
The low-lying SD $2^-$ state is dominantly generated by a p--p type 
excitation of $\nu 1h_{11/2}\otimes \pi 1g_{7/2}$ and $\nu 1g_{9/2}\otimes \pi 1f_{5/2}$ 
in $^{120,122}$Sn and $^{68,70}$Ni, respectively. 
We can see the strongest effect in $^{68}$Ni among these examples. 
This is because the 2qp excitation constructing the SD $2^-$ state is the most p--p type excitation; 
the amplitude $uuuu$ is the largest, giving a large diagonal matrix element. 

\section{Summary}\label{summary}

We have investigated the effect of the pn-pair interaction in the $T=0, S=1$ channel 
on the low-lying spin-dipole (SD) transitions. 
We aimed at clarifying the distinction of the role in between the SD and GT transitions. 
To this end, we have performed a three-body model calculation for the transition $^{80}\mathrm{Ni}\to{}^{80}\mathrm{Cu}$, 
where $^{80}\mathrm{Ni} ={}^{78}\mathrm{Ni}+\mathrm{n}+\mathrm{n}$ and $^{80}\mathrm{Cu} ={}^{78}\mathrm{Ni}+\mathrm{p}+\mathrm{n}$. 
The strength of the $S=1$ pn-pair interaction was varied to see the effect on the SD transition-strength distributions. 
The depth of the mean-field potential was also changed to study the shell effect. 
The effect of the $S=1$ pn-pair interaction depends on the spatial overlap of the pn pair 
and the angular momentum of the valence nucleons. 
The $S=1$ pn-pair interaction in the SD excitations 
is active even if the spatial overlap of the pn pair is weak, 
where the principal quantum numbers of the single-particle orbital are different by up to one unit, 
while only the $\Delta N=0$ excitation is allowed for the GT transition. 
The effect of the $S=1$ pn-pair interaction on the SD transition is thus weaker than on the GT transition. 

To fortify the finding obtained by the three-body model analysis, 
we have performed the nuclear DFT calculations. 
In nuclei where the high-$\ell$ orbitals are located close to the Fermi level, 
we have found a strong effect of the $S=1$ pn-pair interaction. 
In neutron-rich nuclei, the negative-parity states appear in low energies and 
thus the $\beta$-decay rate can be sensitive to the $S=1$ pn-pair interaction, 
similarly for the GT transition. 
A careful analysis of the forbidden $\beta$ decay is thus needed. 

\begin{acknowledgments} 
Discussions with K.~Hagino and H.~Sagawa are acknowledged. 
This work was supported by the JSPS KAKENHI (Grants No. JP19K03824, No. JP19K03861, and No. JP19K03872) 
and the JSPS/NRF/NSFC A3 Foresight Program ``Nuclear Physics in the 21st Century.''
The nuclear DFT calculations were performed on Yukawa-21 
at the Yukawa Institute for Theoretical Physics, Kyoto University.

\end{acknowledgments}

\section*{appendix: angular part of diagonal matrix element}

We summarize here the angular part (\ref{eq:angular}) of the diagonal matrix element of the $S=1$ pair interaction (\ref{eq:spin_trip_int}). 
For the GT operator, we have 
\begin{subequations}
\begin{align}
A^{1^+}(\ell_{j_>},\ell_{j_>};\ell_{j_>},\ell_{j_>})&=
\dfrac{1}{4\pi}\dfrac{(\ell+1)[2(\ell+1)^2+1]}{(2\ell+3)(2\ell+1)}, \\
A^{1^+}(\ell_{j_>},\ell_{j_<};\ell_{j_>},\ell_{j_<})&=
\dfrac{1}{8\pi}\dfrac{3\ell(\ell+1)}{2\ell+1}, \\
A^{1^+}(\ell_{j_<},\ell_{j_<};\ell_{j_<},\ell_{j_<})&=
\dfrac{1}{4\pi}\dfrac{\ell(2\ell^2+1)}{(2\ell+1)(2\ell-1)}.
\end{align}
\end{subequations}

For the SD operators, we have
\begin{align}
A^{0^-}(\ell_{j_<},(\ell-1)_{j_>};\ell_{j_<},(\ell-1)_{j_>})&=\dfrac{\ell}{4\pi}, \label{eq:SD0}
\end{align}
and 
\begin{subequations}
\begin{align}
A^{2^-}(\ell_{j_>},(\ell-1)_{j_>};\ell_{j_>},(\ell-1)_{j_>})& \notag \\
=\dfrac{1}{8\pi}\dfrac{[(2\ell+1)^2+6](\ell+1)\ell}{(2\ell+3)(2\ell+1)(2\ell-1)}, \\
A^{2^-}(\ell_{j_<},(\ell-1)_{j_>};\ell_{j_<},(\ell-1)_{j_>})&=
\dfrac{1}{4\pi}\dfrac{\ell(\ell+1)(\ell-1)}{(2\ell+1)(2\ell-1)}, \\
A^{2^-}(\ell_{j_>},(\ell-1)_{j_<};\ell_{j_>},(\ell-1)_{j_<})&=
\dfrac{1}{8\pi}\dfrac{5\ell(\ell+1)(\ell-1)}{(2\ell+1)(2\ell-1)},  \label{eq:SD2} \\
A^{2^-}(\ell_{j_<},(\ell-1)_{j_<};\ell_{j_<},(\ell-1)_{j_<})& \notag \\
=\dfrac{1}{8\pi}\dfrac{[(2\ell-1)^2+6]\ell(\ell-1)}{(2\ell+1)(2\ell-1)(2\ell-3)} 
\end{align}
\end{subequations}
for $J=0$ and $J=2$, respectively. 

\bibliography{SD_ref}

\end{document}